# Allosteric collaboration between elongation factor G and the ribosomal L1 stalk directs tRNA movements during translation

Jingyi Fei[a], Jonathan E. Bronson[a], Jake M. Hofman[b,1], Rathi L. Srinivas[c], Chris H. Wiggins[d] and Ruben L. Gonzalez, Jr.[a,2]

[a]Department of Chemistry, [b]Department of Physics, [c]The Fu Foundation School of Engineering and Applied Science and [d]Department of Applied Physics and Applied Mathematics, Columbia University, New York, NY 10027

[1]Current address: Yahoo! Research, 111 West 40th Street, 17th Floor, NewYork, NY 10018

[2]To whom correspondence may be addressed. E-mail: rlg2118@columbia.edu

Classification: Biological Sciences, Biochemistry

**ABSTRACT**

Determining the mechanism by which transfer RNAs (tRNAs) rapidly and precisely transit through the ribosomal A, P and E sites during translation remains a major goal in the study of protein synthesis. Here, we report the real-time dynamics of the L1 stalk, a structural element of the large ribosomal subunit that is implicated in directing tRNA movements during translation. Within pre-translocation ribosomal complexes, the L1 stalk exists in a dynamic equilibrium between open and closed conformations. Binding of elongation factor G (EF-G) shifts this equilibrium towards the closed conformation through one of at least two distinct kinetic mechanisms, where the identity of the P-site tRNA dictates the kinetic route that is taken. Within post-translocation complexes, L1 stalk dynamics are dependent on the presence and identity of the E-site tRNA. Collectively, our data demonstrate that EF-G and the L1 stalk allosterically collaborate to direct tRNA translocation from the P to the E sites, and suggest a model for the release of E-site tRNA.

**INTRODUCTION**

During the elongation phase of protein synthesis, the ribosome repetitively cycles through three principal steps: (1) selection of an aminoacyl-transfer RNA (tRNA) at the ribosomal A site (1), (2) peptidyl transfer from the P site-bound peptidyl-tRNA to the A site-bound aminoacyl-tRNA (2) and (3) translocation of the messenger RNA(mRNA)-tRNA complex by one codon, effectively moving the P- and A-site tRNAs into the E- and P-sites, respectively (3). Perhaps the most dynamic steps of this cycle are the precisely directed mRNA and tRNA movements that occur during translocation (3-5). Although this step of the elongation cycle is promoted by elongation factor G (EF-G), numerous biochemical (6), structural (7, 8) and Förster resonance energy transfer (FRET) (9-15) studies have provided strong evidence that the peptidyl transfer step of the elongation cycle spontaneously triggers an EF-G-independent structural rearrangement of the ribosomal pre-translocation (PRE) complex that involves movements of the ribosome-bound tRNAs from their classical- to their hybrid-bound configurations (6-10, 12), movement of the ribosomal L1 stalk from an open to a closed conformation (7, 8, 13, 15) and a counterclockwise, ratchet-like rotation of the small ribosomal subunit relative to the large subunit (7, 8, 11, 14).

Single-molecule FRET (smFRET) investigations have proven a powerful means for directly investigating the conformational dynamics of PRE complexes. Aided by X-ray- and cryogenic electron microscopy (cryo-EM)-derived structural models, several groups have reported kinetic studies of tRNA and ribosome movements within PRE complexes (9, 10, 12-15). tRNA-tRNA smFRET (smFRET$_{tRNA-tRNA}$) experiments initially revealed that upon peptidyl transfer, tRNAs enter into a classical$\rightleftarrows$hybrid dynamic equilibrium within PRE complexes (9, 10, 12). More recently, we have used an L1 stalk-tRNA smFRET (smFRET$_{L1-tRNA}$) signal to demonstrate that upon peptidyl transfer, a direct L1 stalk-tRNA contact is reversibly established (denoted as L1•tRNA) and disrupted (denoted as L1∘tRNA), thus establishing an L1∘tRNA$\rightleftarrows$L1•tRNA dynamic equilibrium within PRE complexes. Using structural arguments, we proposed that L1∘tRNA→L1•tRNA involved a classical→hybrid P-site tRNA transition as well as an open→closed L1 stalk transition and conversely, L1•tRNA→L1∘tRNA involved a hybrid→classical P-site tRNA transition as well as a closed→open L1 stalk transition (13). Furthermore, kinetic analysis of the smFRET$_{tRNA-tRNA}$ and smFRET$_{L1-tRNA}$ signals suggested the possibility that, classical→hybrid and hybrid→classical P-site tRNA transitions might be directly coupled to open→closed and closed→open L1 stalk transitions, respectively, at least at our time resolution (0.05 sec frame$^{-1}$) (13). Unfortunately, at that time the lack of an smFRET signal that could directly and independently report on the open and closed conformations of the L1 stalk precluded direct testing of these hypotheses.

Here we describe a new smFRET signal between ribosomal proteins L1 and L9 (smFRET$_{L1-L9}$) that reports directly on the open and closed conformations of the L1 stalk (Fig. 1). This smFRET$_{L1-L9}$ signal confirms that the L1 stalk indeed fluctuates between open and closed conformations within a PRE complex. Combined with our previous smFRET$_{tRNA-tRNA}$ and smFRET$_{L1-tRNA}$ studies, the data we present here provides strong support for a model in which L1∘tRNA$\rightleftarrows$L1•tRNA fluctuations are composed of coupled classical$\rightleftarrows$hybrid P-site tRNA and open$\rightleftarrows$closed L1 stalk transitions. Since the principal features of the PRE complex L1 stalk dynamics that we report here are in excellent agreement with those reported recently by Cornish *et al.* (15), we will primarily focus on those aspects of L1 stalk dynamics that have not been previously investigated. Specifically, we demonstrate that upon binding to PRE complexes, EF-G allosterically regulates the kinetics of L1 stalk fluctuations, employing one of at least two distinct kinetic strategies in order to shift the equilibrium towards the closed L1 stalk conformation; remarkably, the identity of the P-site tRNA dictates the kinetic strategy used by EF-G. In addition, we report L1 stalk dynamics in post-translocation (POST) complexes and demonstrate that these are dependent on the presence and identity of the E-site tRNA. Based on our results, we propose a role for the L1 stalk in directing the release of deacylated tRNA from the E site.

**RESULTS**

Single-cysteine variants of ribosomal proteins L1 and L9 were fluorescently labeled with Cy5- and Cy3-maleimides, respectively, and reconstituted into large, 50S ribosomal subunits purified from an L1/L9 double-deletion strain of *Escherichia coli* (Fig. 1). Functional testing of dual-labeled 50S subunits using a standard primer extension inhibition assay (16, 17) demonstrated ~90% activity through the first round of translation elongation and ~70% activity in a second round of translation elongation (Methods, Supporting Information (SI) Methods and Fig. S1).

Dual-labeled 50S subunits were used to enzymatically prepare a ribosomal initiation complex (INI) containing fMet-tRNA$^{fMet}$ at the P site (9, 13, 18). Delivery of Phe-tRNA$^{Phe}$, in complex with elongation factor Tu (EF-Tu) and GTP in the presence of EF-G to INI generates a POST complex, POST$_{fM/F}$ (where the subscript denotes the presence of deacylated tRNA$^{fMet}$ in the E site and fMet-Phe-tRNA$^{Phe}$ in the P site) (9, 13, 18).

Using POST$_{fM/F}$ and INI, two PRE complexes were formed. Delivery of EF-Tu(GTP)Lys-tRNA$^{Lys}$ to POST$_{fM/F}$ generates PRE$_{F/K}$ (where the F/K subscript now denotes deacylated tRNA$^{Phe}$ at the P site and fMet-Phe-Lys-tRNA$^{Lys}$ at the A site). Likewise, delivery of EF-Tu(GTP)Phe-tRNA$^{Phe}$ to INI generates PRE$_{fM/F}$, carrying deacylated tRNA$^{fMet}$ at the P site and fMet-Phe-tRNA$^{Phe}$ at the A site (9, 13, 18). Two corresponding PRE-complex analogs were formed by reacting POST$_{fM/F}$ with puromycin to generate PMN$_{F/-}$, containing a deacylated tRNA$^{Phe}$ at the P site and a vacant A site, and reacting INI with puromycin to generate PMN$_{fM/-}$, carrying a deacylated tRNA$^{fMet}$ at the P site and a vacant A site (9, 13, 18).

Analysis of steady-state smFRET vs. time trajectories reveals the presence of three trajectory subpopulations within each of the PRE/PMN complexes (Fig. 2). For all PRE/PMN complexes, the major subpopulation exhibits fluctuations between two well-defined FRET states centered at 0.56 ± 0.01 and 0.34 ± 0.01 FRET (Fig. 2 and 3. Based on close agreement with the smFRET$_{L1-L9}$ values predicted from cryo-EM reconstructions of the open and closed L1 stalk (~0.67 FRET and ~0.35 FRET, assuming $R_0 \approx 55$ Å (19, 20)), the 0.56 and 0.34 FRET states were assigned to the open and closed L1 stalk conformations, respectively. Thus, we will refer to this trajectory subpopulation as SP$_{fluct}$. The remaining two subpopulations exhibit either stable 0.56 FRET (SP$_{open}$) or stable 0.34 FRET (SP$_{closed}$) prior to fluorophore photobleaching (Figs. 2 and 3). SP$_{open}$ is attributed to: (1) contaminating amounts of POST$_{fM/F}$ or INI that failed to react with EF-Tu(GTP)aminoacyl-tRNA or puromycin and (2) PRE/PMN complexes that exhibited photobleaching directly out of the open conformation prior to undergoing a open→closed L1 stalk transition. SP$_{closed}$ is attributed to PRE/PMN complexes that exhibited photobleaching directly out of the closed conformation prior to undergoing a closed→open L1 stalk transition. We note here that PRE/PMN trajectories that occupy SP$_{open}$ and SP$_{closed}$ do not correspond to static subpopulations of PRE/PMN complexes that are somehow distinct from the fluctuating subpopulation of PRE/PMN complexes. Rather, the occupancies of SP$_{fluct}$, SP$_{open}$ and SP$_{closed}$ are simply determined by a competition between open⇌closed L1 stalk transitions and photobleaching from the open and closed states (Methods and refs. (18, 21).

Fluctuations within SP$_{fluct}$ for all PRE/PMN complexes occur within one frame (Fig. 2), suggesting that open⇌closed L1 stalk transitions occur without sampling any intermediate state(s), at least not within our time resolution (0.10 sec frame$^{-1}$, see SI Methods for information regarding the time resolution of the smFRET data). Consistent with this, transition density plots reveal the existence of two major L1 stalk transitions, open→closed and closed→open, with no evidence of any significantly populated intermediate state(s) (Fig. S2).

Rates for L1 stalk closing and opening ($k_{close}$ and $k_{open}$) were extracted using dwell time analyses of $SP_{fluct}$ for all PRE/PMN complexes (Methods, Fig. S2 and Table 1). In addition, rates for the formation and disruption of the L1 stalk-tRNA contact ($k_{L1 \cdot tRNA}$ and $k_{L1 \circ tRNA}$) have been previously reported for PRE complexes analogous to $PRE_{F/K}$ and $PMN_{F/-}$ (Ref. (13) and Table 1) and are measured and reported here for $PRE_{fM/F}$ and $PMN_{fM/-}$ (Fig. S3 and Table 1). Table 1 demonstrates the close agreement between $k_{close}$ and $k_{L1 \cdot tRNA}$ and between $k_{open}$ and $k_{L1 \circ tRNA}$ for all PRE/PMN complexes. Significantly, $k_{close}$ and $k_{open}$ exhibit a dependence on the presence of an A-site peptidyl-tRNA that very closely mirrors the dependence observed for $k_{L1 \cdot tRNA}$ and $k_{L1 \circ tRNA}$ (compare changes in $k_{close}$ to those in $k_{L1 \cdot tRNA}$ and changes in $k_{open}$ to those in $k_{L1 \circ tRNA}$ for $PRE_{F/K}$ vs. $PMN_{F/-}$ and $PRE_{fM/F}$ vs. $PMN_{fM/-}$). The slight discrepancy between $k_{close}$ and $k_{L1 \cdot tRNA}$ in $PRE_{fM/F}$ vs. $PMN_{fM/-}$ most likely originates from the presence of the Cy3 fluorophore on tRNA$^{fMet}$ in the $k_{L1 \cdot tRNA}$ measurement (note that Cy3 on tRNA$^{fMet}$ is at a different position than on tRNA$^{Phe}$) or, less likely, suggests that coupling between closing of the L1 stalk and movement of tRNA into the hybrid configuration might depend on the identity of P-site tRNA. Likewise, Table 1 demonstrates that $k_{close}$ and $k_{open}$ exhibit a dependence on the identity of the P-site tRNA that very closely mirrors the dependence observed for $k_{L1 \cdot tRNA}$ and $k_{L1 \circ tRNA}$ (compare changes in $k_{close}$ to those in $k_{L1 \cdot tRNA}$ and changes in $k_{open}$ to those in $k_{L1 \circ tRNA}$ for $PRE_{fM/F}$ vs. $PRE_{F/K}$ and $PMN_{fM/-}$ vs. $PMN_{F/-}$). Here our observations are consistent with the well-documented propensity of tRNA$^{fMet}$ to occupy the classical configuration (11, 22, 23). Collectively, our data provide strong support for the tight coupling of open→closed L1 stalk and classical→hybrid tRNA transitions on the one hand and closed→open L1 stalk and hybrid→classical tRNA transitions on the other.

We have previously shown that addition of 1 µM EF-G in the presence of GDPNP (a non-hydrolyzable GTP analog) to a PMN complex analogous to $PMN_{F/-}$ significantly inhibits L1•tRNA→L1∘tRNA transitions, thereby shifting the L1∘tRNA ⇌ L1•tRNA equilibrium strongly towards L1•tRNA (13). In a completely analogous manner, addition of 1 µM EF-G(GDPNP) to $PMN_{F/-}$ strongly inhibits closed→open L1 stalk transitions such that the open ⇌ closed L1 stalk equilibrium shifts to favor the closed L1 stalk conformation and the rate of photobleaching from the closed L1 stalk conformation effectively out competes closed→open transitions; the overall effect is a decrease in the occupancy of $SP_{fluct}$ and a corresponding increase in the occupancy of $SP_{closed}$ (Fig. 2, 4A and S4). Because EF-G(GDPNP)-bound $PMN_{F/-}$ occupies $SP_{closed}$, $k_{open}$ and $k_{close}$ for EF-G(GDPNP)-bound $PMN_{F/-}$ cannot be calculated (Table 1). Nevertheless, it is clear that without directly contacting either the P-site tRNA or the L1 stalk, binding of EF-G(GDPNP) to $PMN_{F/-}$ suppresses both L1•tRNA→L1∘tRNA and closed→open L1 stalk transitions; this observation strongly suggests that during translocation EF-G-ribosome interactions allosterically regulate tRNA as well as L1 stalk dynamics.

Addition of 1 µM EF-G(GDPNP) to $PMN_{fM/-}$ has a dramatically different effect than that observed for $PMN_{F/-}$. Rather than shifting the trajectory subpopulation occupancy towards $SP_{closed}$, binding of EF-G(GDPNP) to $PMN_{fM/-}$ leads to preferential occupancy of $SP_{fluct}$ (Fig. 2 and S4). However, contour plots of the time evolution of population FRET reveal that, like $PMN_{F/-}$, $PMN_{fM/-}$ preferentially occupies the closed conformation of the L1 stalk in the presence of EF-G(GDPNP) (Fig. 4). In order to investigate the kinetic basis for the preferential occupancy of the closed L1 stalk conformation, we determined $k_{close}$ and $k_{open}$ for $PMN_{fM/-}$ in the absence and presence of 1 μM EF-G(GDPNP) (Tables 1 and S2). The data in Table 1 demonstrate that EF-G(GDPNP) primarily increases $k_{close}$ by ~8-fold and has only a relatively minor effect on $k_{open}$. Thus, rather than suppressing closed→open L1 stalk transitions, as was observed for $PMN_{F/-}$, EF-G(GDPNP) increases $k_{close}$ by destabilizing the open conformation of the L1 stalk in $PMN_{fM/-}$, resulting in an overall shift of the open⇌closed equilibrium analogous to what is observed for EF-G(GDPNP) binding to $PMN_{F/-}$ (i.e. the equilibrium shifts to favor the closed L1 stalk conformation). Here, the increased $k_{close}$ and unchanged $k_{open}$ yield a decrease in the occupancy of $SP_{open}$ and a corresponding increase in the occupancy of $SP_{fluct}$ and, to a lesser extent, $SP_{closed}$ (Fig. 2 and 4). These results reveal that, although the overall effect of EF-G binding to PRE complexes is to shift the open⇌closed L1 stalk equilibrium towards the closed L1 stalk conformation, distinct kinetic mechanisms that depend on the identity of the P-site tRNA are used in order to accomplish this. Fully consistent with these results, new $smFRET_{L1-tRNA}$ experiments on a PMN complex analogous to $PMN_{fM/-}$ reveal that in the presence of 1 μM EF-G(GDPNP), the majority of $smFRET_{L1-tRNA}$ trajectories fluctuate between L1∘tRNA and L1•tRNA, with a preference for L1•tRNA that is primarily driven by an ~4-fold increase in $k_{L1•tRNA}$ (Fig. S3 and Table 1).

In addition to characterization of L1 stalk dynamics within PRE/PMN complexes, the $smFRET_{L1-L9}$ signal allows investigation of L1 stalk dynamics within POST complexes (Fig. 5). The majority of $POST_{fM/F}$ trajectories occupy $SP_{open}$, indicating a strong preference for the open L1 stalk conformation (Fig. 2 and 5A). Within $SP_{fluct}$, $k_{close}$ = 0.10 ± 0.13 $sec^{-1}$ and $k_{open}$ = 0.99 ± 0.16 $sec^{-1}$, also yielding a preference for occupying the open L1 stalk conformation. Because of heterogeneity in the tRNA occupancy of the E site, we generated a homogenous $POST_{-/F}$ complex by quantitatively dissociating $tRNA^{fMet}$ from $POST_{fM/F}$ (refs. (24, 25) and Fig. S5). Figure S6 shows that quantitative dissociation of the E-site tRNA decreases the occupancy of $SP_{fluct}$ in favor of $SP_{open}$, strengthening the preference of the POST complex for the open L1 stalk conformation.

To test the generality of these results, we repeated these experiments on $POST_{F/K}$. In contrast to $POST_{fM/F}$, we find that only a minority of $POST_{F/K}$ trajectories occupy $SP_{open}$. Instead, the majority of $POST_{F/K}$ trajectories occupy $SP_{fluct}$, with $k_{close}$ = 0.31 ± 0.09 $sec^{-1}$ and $k_{open}$ = 0.76 ± 0.22 $sec^{-1}$, again generating a preference for the open L1 stalk conformation (Fig. 2 and 5B). Because heterogeneity in the E-site tRNA occupancy of $POST_{fM/F}$ and $POST_{F/K}$ is similar (Fig. S5), our data suggest that L1 stalk dynamics in POST complexes may depend on the identity of the E-site tRNA. Quantitative dissociation of deacylated $tRNA^{Phe}$ from the E site of $POST_{F/K}$ (Fig. S5) reveals that the majority of $POST_{-/K}$ trajectories occupy $SP_{open}$, completely analogous to our observations on $POST_{-/F}$ (Fig. S6). Thus, although the L1 stalk within POST complexes exhibits an overall preference for the open conformation, the kinetics underlying this preference depend on the presence and identity of the E-site tRNA. This observation implies that each tRNA species might make slightly different and unique binding interactions with the ribosomal E site.

In order to assess the dynamics of the L1 stalk within a homogeneous POST

complex containing a fully occupied E site, we artificially delivered 1 μM deacylated tRNA$^{fMet}$ to POST$_{-/F}$ to generate POST$_{fM*/F}$ and deacylated tRNA$^{Phe}$ to POST$_{-/K}$ to generate POST$_{F*/K}$. In stark contrast to our results for POST$_{fM/F}$ and POST$_{F/K}$, containing authentically-translocated E-site tRNAs, we find that the majority of POST$_{fM*/F}$ and POST$_{F*/K}$ trajectories preferentially occupy the closed L1 stalk conformation (compare Fig. 5A to Fig. S6D and Fig. 5B to Fig. S6E). Preliminary subpopulation and kinetic analysis of POST$_{fM*/F}$ and POST$_{F*/K}$ (Fig. S6) indicate that, similar to our results for POST complexes containing authentically-translocated E-site tRNAs, L1 stalk dynamics in POST complexes containing artificially-delivered E-site tRNAs may also depend on the identity of the E-site tRNA. It should be stated, however, that possible compositional heterogeneity arising from the incomplete binding of deacylated tRNA to the ribosomal E site and/or from reverse translocation of POST$_{fM*/F}$ and POST$_{F*/K}$ (these experiments were conducted in the absence of EF-G) (26, 27) precludes detailed subpopulation and kinetic analysis. Regardless, it is clear that while authentically-translocated E-site tRNAs exhibit a strong preference for the open conformation of the L1 stalk, artificially-delivered E-site tRNAs instead generate a preference for the closed L1 stalk conformation. Presumably, this closed L1 stalk conformation is identical to the "half-closed" conformation that has been observed by Cornish *et al.* in similarly prepared POST complexes (i.e. containing an artificially-delivered E-site tRNA) (15). Thus, it seems that our smFRET$_{L1-L9}$ signal cannot distinguish between the fully-closed L1 stalk conformation observed in PRE/PMN complexes and the half-closed L1 stalk conformation observed in POST complexes containing an artificially-delivered E-site tRNA; this is perhaps not surprising given the smaller dynamic range of the smFRET$_{L1-L9}$ signal in this work relative to the L1-L33 smFRET signal in Cornish *et al* (15) Thus, whether or not the half-closed L1 stalk conformation observed by Cornish *et al.* in POST complexes containing artificially-delivered E-site tRNAs is sampled in POST complexes containing authentically-translocated E-site tRNAs remains to be determined. Regardless, our results demonstrate that L1 stalk dynamics within POST complexes are sensitive to the mechanism through which the deacylated tRNA enters the E site.

**DISCUSSION**

Previously we have proposed that PRE/PMN complexes spontaneously and reversibly fluctuate between two major conformational states: global state 1 (GS1), encompassing classically-bound tRNAs, an open L1 stalk and a non-ratcheted ribosome and global state 2 (GS2), encompassing hybrid-bound tRNAs, a closed L1 stalk and a ratcheted ribosome (13). Consistent with this model, our smFRET$_{L1-L9}$ results demonstrate that the L1 stalk within PRE/PMN complexes exists in an open$\rightleftarrows$closed dynamic equilibrium which exhibits kinetics closely matching those of the classical$\rightleftarrows$hybrid tRNA (9) and the L1∘tRNA$\rightleftarrows$L1•tRNA (13) dynamic equilibria. Most notably, all three equilibria have matching kinetic responses towards the occupancy of the A site by a peptidyl-tRNA and the identity of the P-site tRNA. These kinetic data demonstrate the close coupling between tRNA and L1 stalk dynamics within PRE/PMN complexes. In addition to our smFRET$_{tRNA-tRNA}$ (9), smFRET$_{L1-tRNA}$ (13) and smFRET$_{L1-L9}$ studies, further support for the GS1$\rightleftarrows$GS2 model is provided by two recent smFRET studies from Ha, Noller and coworkers demonstrating the close correlation between the equilibrium constants governing the non-ratcheted$\rightleftarrows$ratcheted ribosome and open$\rightleftarrows$closed L1 stalk equilibria (14, 15). In complete agreement with the smFRET results and the GS1$\rightleftarrows$GS2 model, two recent cryo-EM studies applied particle classification methods to reveal the existence of two major PRE complex conformations within a single pre-translocation sample, corresponding to GS1 and GS2, respectively (7, 8). Despite the excellent agreement between the GS1⇌GS2 model and the available smFRET and cryo-EM data, however, it remains possible that short-lived and/or rarely sampled intermediates within GS1→GS2 and/or GS2→GS1 transitions have thus far eluded detection by smFRET experiments or cryo-EM reconstructions. Future smFRET experiments recorded at higher-time resolution or using ribosome-targeting small-molecule translocation inhibitors or mutagenized ribosomes may prove useful tools for uncovering such short-lived and/or rarely-sampled intermediates. Nevertheless, the GS1⇌GS2 model represents a simple dynamic model which is consistent with the available data and provides a convenient framework for describing the global dynamics of the translating ribosome.

Binding of EF-G(GDPNP) to PMN complexes strongly shifts the open⇌closed L1 stalk equilibrium towards the closed conformation by regulating $k_{open}$ and/or $k_{closed}$. Given that EF-G(GDPNP) binds near the A site of the PRE complex, ~170 Å away from the hinge region of the L1 stalk (28-31), our data demonstrates that EF-G(GDPNP) regulates L1 stalk dynamics allosterically, through its interactions with the ribosome upon binding to the PMN complex. An attractive hypothesis, consistent with the close coupling of ratcheting, L1 stalk and tRNA dynamics stipulated by the GS1⇌GS2 model, is that EF-G(GDPNP) establishes interactions with the ribosome that directly stabilize the ratcheted conformation of the ribosome, indirectly leading to stabilization of the hybrid P-site tRNA configuration and the closed L1 stalk conformation. Indeed, the discovery that vacant *Saccharomyces cerevisiae* ribosomes (i.e. not containing tRNA substrates) predominantly exist in a ratcheted conformation with a closed L1 stalk (32) suggests the possibility that the coupling between intersubunit ratcheting and L1 stalk closure might be independent of the presence of a P-site tRNA and may instead be encoded within the architecture of the ribosome itself. Evidence for a similar possibility in prokaryotic ribosomes comes from the correlation between the equilibrium constants governing the non-ratcheted⇌ratcheted ribosome and open⇌closed L1 stalk equilibria in vacant *E. coli* ribosomes (14, 15); the correlation between the forward and reverse rates of these two processes, however, remains to be investigated. Regardless, in this framework, ratcheting and L1 stalk closure would function allosterically in order to promote and stabilize the hybrid tRNA configuration during translocation. Future experiments exploring the role of ribosomal structural elements in regulating ratcheting, L1 stalk and tRNA dynamics should allow testing of these hypotheses.

Using $smFRET_{L1-tRNA}$ and $smFRET_{L1-L9}$ signals as reporters for the GS1⇌GS2 equilibrium, we find that EF-G(GDPNP) can shift the GS1⇌GS2 equilibrium towards GS2 through at least two distinct kinetic mechanisms, the choice of which is regulated by the identity of the P-site tRNA. When $tRNA^{Phe}$ occupies the P site, EF-G(GDPNP) almost completely suppresses GS2→GS1 transitions whereas when $tRNA^{fMet}$ occupies the P site, EF-G(GDPNP) has an almost negligible effect on GS2→GS1 transitions, instead increasing the rate of GS1→GS2 transitions by ~4-8-fold. This latter result strongly suggests that EF-G can bind directly to PRE complexes in the GS1 state and actively promote the GS1→GS2 transition, although the extent to which this is observed strongly depends on the identity of the P-site tRNA.

The conformational dynamics of the L1 stalk observed within POST

complexes, which are likely uncoupled from intersubunit ratcheting dynamics since POST complexes are not expected to ratchet (11, 14, 30), are altogether distinct from those observed within PRE/PMN complexes. We find that L1 stalk dynamics within POST complexes are sensitive to the presence and identity of the E-site tRNA as well as to the mechanism through which the tRNA enters the E site. In the presence of a vacant E site, the L1 stalk is almost uniformly found in the stable open conformation. The presence of an authentically-translocated E-site tRNA, however, triggers open$\rightleftarrows$closed L1 stalk fluctuations where $k_{open}$ and $k_{close}$ depend on the identity of the E-site tRNA. Despite the differing kinetics, POST complexes containing either authentically-translocated tRNA$^{fMet}$ or tRNA$^{Phe}$ both favor the open L1 stalk conformation; this stands in contrast with the uniformly half-closed L1 stalk conformation observed within POST complexes containing an artificially-delivered E-site tRNA (15) and in X-ray crystal structures of POST-like ribosomes carrying what are likely artificially-delivered E-site tRNAs (33-35). Consistent with these observations, we find that artificial delivery of a deacylated tRNA into the E site of a POST complex triggers a strong preference for the closed L1 stalk conformation.

Previously we have reported a stable (i.e. non-fluctuating) high FRET signal between the L1 stalk and an authentically-translocated E-site tRNA within a POST complex (13). In contrast, the data we present here demonstrates that the L1 stalk within an analogous POST complex, under identical sample conditions as our previous study, undergoes open$\rightleftarrows$closed fluctuations. In order to reconcile these two observations, we propose a model in which the authentically-translocated E-site tRNA is reconfigured within the E site such that the direct interaction between the tRNA and the L1 stalk is maintained during the open$\rightleftarrows$closed fluctuations of the L1 stalk. This model is strongly supported by the observation that the E-site tRNA occupies one configuration in X-ray crystal structures of POST-like complexes bearing a closed, or half-closed, L1 stalk conformation (33-35) but occupies a notably different E-site tRNA configuration in cryo-EM reconstructions of POST complexes bearing an open L1 stalk conformation(30). As originally suggested by the authors of the cryo-EM study (30), reconfiguration of the E-site tRNA such that a direct contact with the opening L1 stalk is maintained may be mechanistically important for E-site tRNA release. That said, we find that $k_{open}$ is ~10-fold faster than the rate of passive tRNA dissociation from the E site in POST complexes, indicating that the L1 stalk/E-site tRNA can undergo numerous fluctuations before the E-site tRNA dissociates and strongly suggesting that opening of the L1 stalk is not rate limiting for E-site tRNA release. Finally, the observation that L1 stalk dynamics within a POST complex depend on the identity of the E-site tRNA may reflect a difference in the energetics of reconfiguring each tRNA species within the E site. The molecular basis for this difference likely originates from the slightly different interactions that each tRNA would be expected to make with structural elements of the ribosomal E site.

Collectively, our data demonstrate that differences in the interactions of tRNA$^{fMet}$ and tRNA$^{Phe}$ with the elongating ribosome can: (1) bias the kinetics of GS1$\rightleftarrows$GS2 transitions in PRE/PMN complexes; (2) control the kinetic mechanism through which EF-G stabilizes the GS2 state during translocation; and (3) regulate tRNA and L1 stalk dynamics within POST complexes. It remains to be investigated whether these differences are due to tRNA identity elements that uniquely distinguish initiator tRNA$^{fMet}$ from all elongator tRNAs (36), thus suggesting that elongator tRNAs will generally exhibit kinetic behavior similar to tRNA$^{Phe}$, or whether similarly significant differences in kinetic behavior will be found even among elongator tRNAs. Future smFRET studies using an expanded set of elongator tRNAs and/or tRNA$^{fMet}$ variants containing mutations to tRNA$^{fMet}$ identity elements, should reveal which features of tRNA structure and tRNA-ribosome interactions are involved in regulating the kinetic behavior of PRE/PMN and POST complexes.

## METHODS SUMMARY

All experiments were performed in Tris-polymix buffer (50mM Tris-OAc, 100mM KCl, 5mM NH$_4$OAc, 0.5mM Ca(OAc)$_2$, 10mM 2-mercaptoethanol, 5mM putrescine and 1mM spermidine) at 15mM Mg(OAc)$_2$ and at pH$_{25°C}$=7.5 (9). smFRET trajectories were recorded using a home-built total internal reflection fluorescence microscope (13, 18). Each smFRET trajectory was idealized as a hidden Markov model, using the vbFRET software package (37). Dwell times spent in each state prior to transitioning were extracted from the idealized smFRET trajectories and the lifetime of each state was determined by exponential fitting of the corresponding one-dimensional population *vs.* time histogram (9, 13, 18). Transition rates were calculated by taking the inverse of the lifetimes and correcting for the rate of photobleaching from each state. Full methods and references can be found in the SI Methods.

## ACKNOWLEDGEMENTS

This work was supported by grants to R.L.G. from the Burroughs Wellcome Fund (CABS 1004856), the NSF (MCB 0644262) and the NIH-NIGMS (1RO1GM084288-01). C.H.W. was supported by the NIH (1U54CA121852-01A1 and 5PN2EY016586-03). R.L.S. was supported by an NSF-REU. We thank S. Das for managing the Gonzalez laboratory, J. Frank, H. Gao and X. Agirrezabala for providing us with GS1 and GS2 structural models, and M.M. Elvekrog, D.D. MacDougall and S.H. Sternberg for valuable discussions and comments on the manuscript.

## AUTHOR CONTRIBUTIONS

J.F. and R.L.G. designed the research; J.F. conducted the research; J.B., J.M.H and C.H.W. developed vbFRET; J.F. and J.B. analyzed the data; R.L.S. assisted J.F. with cloning, mutagenesis and purification of L9; J.F. and R.L.G. wrote the manuscript; all authors approved the final manuscript.

**Table 1**. Transition rates for L1 stalk closing ($k_{close}$) and opening ($k_{open}$), as well as the formation ($k_{L1 \bullet tRNA}$) and disruption ($k_{L1 \circ tRNA}$) of the L1-tRNA interaction for PRE/PMN complexes [a].

| Complex | $k_{close}$ (sec$^{-1}$) | $k_{open}$ (sec$^{-1}$) | $k_{L1 \bullet tRNA}$ (sec$^{-1}$) | $k_{L1 \circ tRNA}$ (sec$^{-1}$) |
|---|---|---|---|---|
| PRE$_{F/K}$ | $k_1$ = 3.4±0.4 (A$_1$ = 72±8 %) <br> $k_2$ = 0.45±0.12 (A$_2$ = 28±8 %)[b] | 0.75±0.14 | $k_1$ = 2.9±0.2 (A$_1$ = 70±3 %)[c] <br> $k_2$ = 0.32±0.05 (A$_2$ = 30±3 %)[b,c] | 0.85±0.04[c] |
| PMN$_{F/-}$ | 0.51±0.06 | 0.84±0.17 | 0.43±0.03[c] | 1.06±0.05[c] |
| PRE$_{fM/F}$ | $k_1$ = 2.0±0.6 (A$_1$ = 60±2 %) <br> $k_2$ = 0.43±0.06 (A$_2$ = 40±3 %)[b] | 1.8±0.3 | $k_1$ = 2.8±0.2 (A$_1$ = 73±4 %) <br> $k_2$ = 0.69±0.04 (A$_2$ = 27±4 %)[b] | 3.0±0.4 |
| PMN$_{fM/-}$ | 0.37±0.09 | 1.5±0.3 | 0.36±0.11 | 2.6±0.2 |
| PMN$_{F/-}$ + EF-G(GDPNP)[d] | – | – | – | – |
| PMN$_{fM/-}$ + EF-G(GDPNP) | 3.0±0.6 | 1.2±0.2 | 1.4±0.1 | 1.54±0.04 |

[a]  Rates reported here are the average and standard deviation from three or four independent data sets. All rates were corrected for photobleaching (see Methods and Table S1).

[b]  The dwell time histograms for the open L1 stalk conformation and the disrupted L1 stalk-tRNA interaction in PRE$_{F/K}$ and PRE$_{fM/F}$ were better described by a double-exponential decay. The ~30% population with the slower rate was assigned to complexes in which the peptidyl-tRNA has dissociated from the A site (9, 13, 38).

[c]  Rates for the formation (k$_{L1 \bullet tRNA}$) and disruption (k$_{L1 \circ tRNA}$) of the L1-tRNA interaction in PRE$_{F/K}$ and PMN$_{F/-}$ were reanalyzed using vbFRET (see Methods) and the previously recorded raw data (13). Average values and standard deviations were calculated the as previously reported (13).

[d]  The major effect of EF-G(GDPNP) binding to PMN$_{F/-}$ is to shift the trajectory subpopulation occupancy towards SP$_{closed}$; thus, $k_{open}$ and $k_{close}$ for the EF-G(GDPNP)-bound fraction of PMN$_{F/-}$ cannot be calculated. Despite this, incomplete reactivity at each of the various enzymatic steps required to prepare PMN$_{F/-}$ yields a residual amount of partially-reacted complexes which result in trajectories that occupy SP$_{fluct}$ (see Fig. 2B). For details regarding this compositional heterogeneity of the PMN$_{F/-}$ sample and a detailed kinetic analysis of SP$_{fluct}$ for this sample please see SI Methods, Fig S4, and Table S2.

## FIGURES

*Figure 1.* **Fluorescent labeling of ribosomal proteins L1 and L9 within the 50S ribosomal subunit.** X-ray crystallographic structure of the 50S subunit (PDB ID 2J01). The FRET donor (Cy3, green star) and acceptor (Cy5, red star) fluorophores are denoted at approximate positions on ribosomal protein L9 (cyan) and L1 (dark blue).

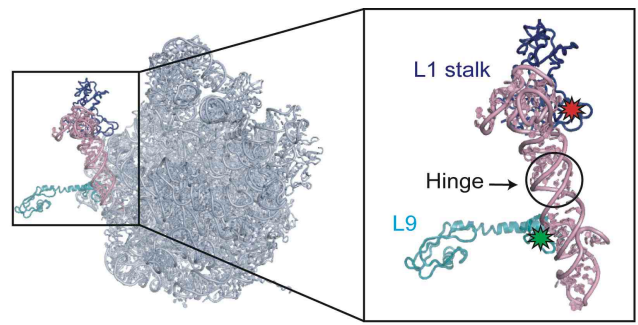

*Figure 2.* **Sample smFRET *vs.* time trajectories and relative occupancies of trajectory subpopulations. (A)** Three subpopulations of trajectories were identified: stable 0.56 FRET (SP$_{open}$, left panel), fluctuating between 0.56 and 0.34 FRET (SP$_{fluct}$, middle panel), and stable 0.34 FRET (SP$_{closed}$, right panel). Representative Cy3 and Cy5 emission intensities are shown in green and red, respectively (top row). The corresponding smFRET traces, $I_{Cy5}/(I_{Cy3}+I_{Cy5})$, are shown in blue (bottom row). **(B)** The percentage of trajectories occupying SP$_{open}$, SP$_{fluct}$ and SP$_{closed}$ are shown as bar graphs for each complex. The mean and the standard deviation of the occupancy for each subpopulation in each complex, shown in red numbers, was calculated from four independent data sets.

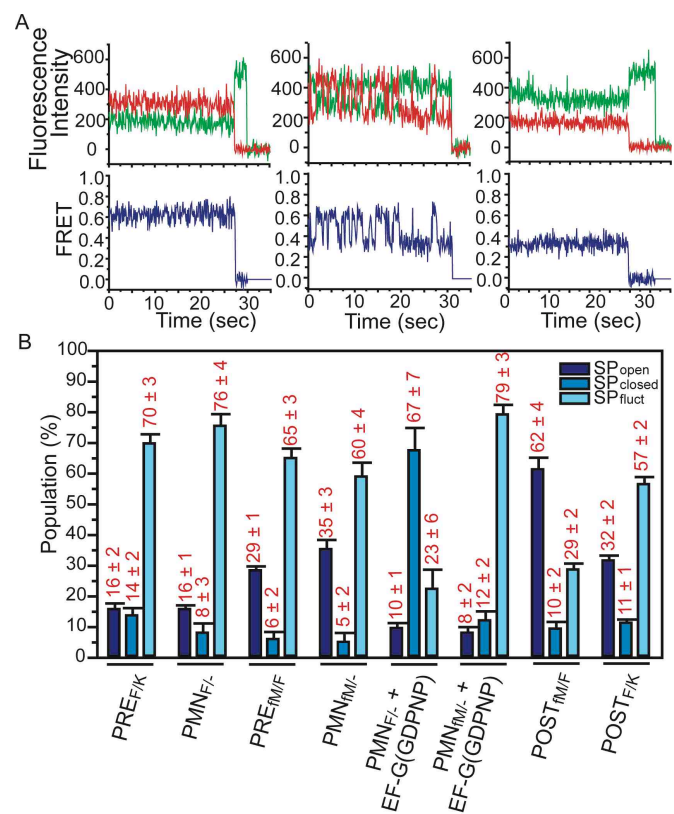

*Figure 3.* **The L1 stalk fluctuates between open and closed conformations in PRE/PMN complexes.** Cartoon representations of the various complexes depict the 30S and 50S subunits in tan and lavender, respectively, with the L1 stalk in dark blue, L9 in cyan and tRNA$^{fMet}$, tRNA$^{Phe}$ and tRNA$^{Lys}$ as orange, brown and purple lines, respectively. Surface contour plots of the time evolution of population FRET are plotted from tan (lowest population) to red (highest population). The number of traces that were used to construct each surface contour plot is indicated by "N." (A) PRE$_{F/K}$ was generated by addition of 100 nM EF-Tu(GTP)Lys-tRNA$^{Lys}$ to POST$_{fM/F}$. (B) PMN$_{F/-}$ was generated by addition of 1 mM puromycin to POST$_{fM/F}$. (C) PRE$_{fM/F}$ was generated by addition of 100 nM EF-Tu(GTP)Phe-tRNA$^{Phe}$ to INI. (D) PMN$_{fM/-}$ was generated by addition of 1 mM puromycin to INI.

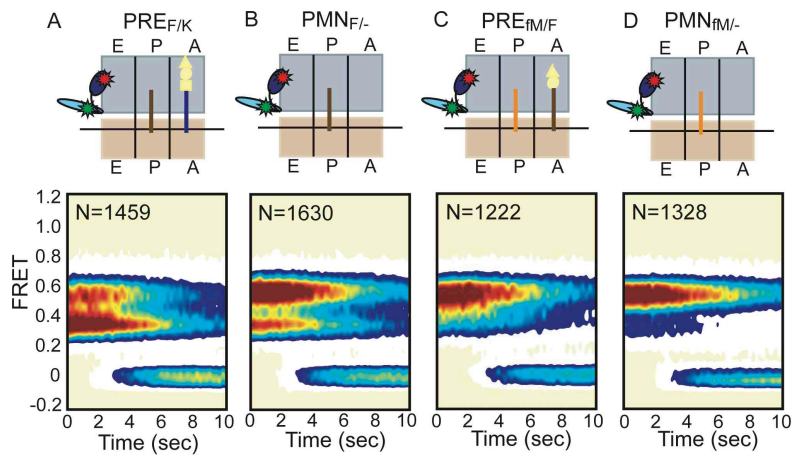

*Figure 4* **EF-G allosterically regulates L1 stalk dynamics in a P-site tRNA dependent manner.** Data are displayed as in Figure 3. 1 μM EF-G(GDPNP) was added to (A) PMN$_{F/-}$ and (B) PMN$_{fM/-}$.

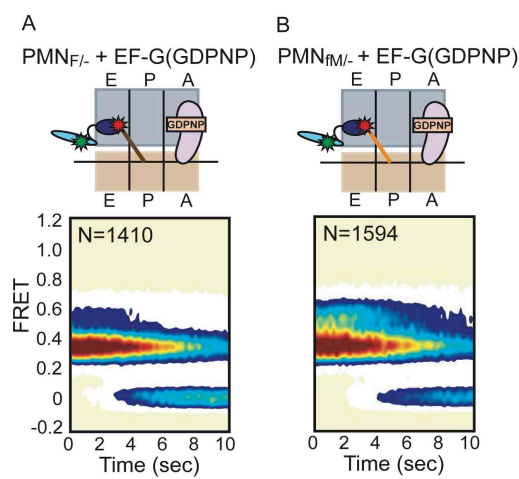

*Figure 5* **The L1 stalk undergoes conformational dynamics within POST complexes.**

Data are displayed as in Figure 3. (A) POST$_{fM/F}$. (B) POST$_{F/K}$ was generated by addition of 100 nM EF-Tu(GTP)Lys-tRNA$^{Lys}$ and 1 μM EF-G(GTP) to POST$_{fM/F}$.

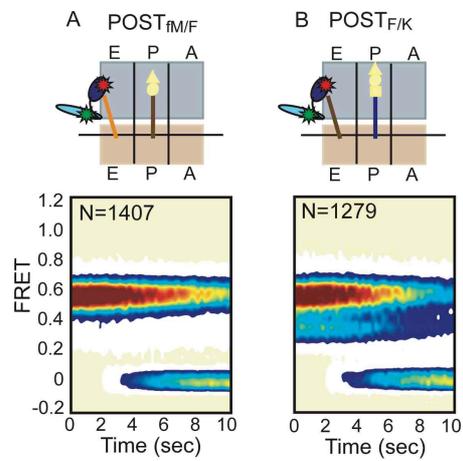

# Allosteric collaboration between elongation factor G and the ribosomal L1 stalk direct tRNA movements during translation

Jingyi Fei[a], Jonathan E. Bronson[a], Jake M. Hofman[b,1], Rathi L. Srinivas[c], Chris H. Wiggins[d] and Ruben L. Gonzalez, Jr.[a,2]

[a]Department of Chemistry, [b]Department of Physics, [c]The Fu Foundation School of Engineering and Applied Science and [d]Department of Applied Physics and Applied Mathematics, Columbia University, New York, NY 10027

[1]Current address: Yahoo! Research, 111 West 40th Street, 17th Floor, NewYork, NY 10018

[2]To whom correspondence may be addressed. E-mail: rlg2118@columbia.edu

Classification: Biological Sciences, Biochemistry

## Supporting Methods

***Buffer conditions.*** Biochemical experiments were performed in Tris-polymix buffer (50mM Tris-OAc, 100mM KCl, 5mM $NH_4OAc$, 0.5mM $Ca(OAc)_2$, 10mM 2-mercaptoethanol, 5mM putrescine, and 1mM spermidine) at 15mM $Mg(OAc)_2$ and at $pH_{25°C}$=7.5. Single-molecule experiments were conducted in an identical buffer, supplemented with an oxygen-scavenging system (300 μg/mL glucose oxidase, 40 μg/mL catalase and 1% β-D-glucose) (1, 2) and a triplet-state quenching cocktail (1mM 1,3,5,7-cyclooctatetraene (Aldrich) and 1mM *p*-nitrobenzyl alcohol (Fluka)) (3).

***Preparation of translation factors, tRNAs, and mRNA.*** All translation factors were purified as previously reported (1). $tRNA^{fMet}$ was labeled with Cy3-maleimide at the $s^4U8$ position (1, 4). $tRNA^{fMet}$, $(Cy3)tRNA^{fMet}$, $tRNA^{Phe}$, and $tRNA^{Lys}$ were aminoacylated with the corresponding amino acids, and $Met-tRNA^{fMet}$ and $Met-(Cy3)tRNA^{fMet}$ were formylated, as previously described (1). A T4 gene product 32-derived mRNA was chemically synthesized (Dharmacon, Inc) to contain a 5'-biotin followed by an 18 nucleotide spacer, a strong Shine-Dalgarno (AAAGGA) sequence, nucleotides encoding fMet, Phe, Lys, as the first three amino acids, and an additional six amino acids.

***E. coli L1/L9 double deletion strain.*** L1 and L9 single deletion strains of *E. coli* were generated from a wild-type *E. coli* strain using the one-step technique reported by Datsenko and Wanner (5, 6). The original L1 and L9 genes in the wild-type *E. coli* strain were replaced by kanamycin and chloramphenicol resistance cassettes, respectively. The L1/L9 double deletion strain was subsequently generated using P1*vir* phage transduction from the single deletion strains following a protocol provided by Prof. Robert T. Sauer (Department of Biology, Massachusetts Institute of Technology) (http://openwetware.org/wiki/Sauer:P1vir_phage_transduction) (7). First the L9 single deletion strain was infected with P1*vir* phage. The resulting P1*vir* phage lysate, containing transducing particles carrying random sections of the L9 single deletion strain genome, including the chloramphenicol resistance cassette located at the former position of the gene encoding L9, was then used to infect a liquid culture of the L1 single deletion strain. The P1*vir* phage infected culture of the L1 single deletion strain was plated and colonies exhibiting both kanamycin and chloramphenicol resistance were selected. Deletion of both the L1 and L9 genes was verified by PCR amplification and DNA sequencing. L1 and L9 double deletion strains exhibit a slow-growth phenotype, with a doubling rate that is ~6-7 fold slower than wild-type *E. coli*.

**Purification of 50S subunits lacking L1 and L9.** 50S subunits lacking L1 and L9 were purified from the L1/L9 double deletion strain of *E. coli* by sucrose density gradient ultracentrifugation using a previously described purification protocol (1, 8). Two-dimensional SDS-PAGE was used to verify the absence of L1 and L9 from the purified subunits.

**Design and construction of fluorescently-labeled L1 and L9 mutants.** The genes encoding *E. coli* L1 and L9 were cloned from C600 genomic DNA into the pProEX-HTb plasmid system, which contains an N-terminal six-histidine (6xHis) affinity purification tag separated from the cloned gene by a tobacco etch virus (TeV) protease cleavage site (1, 8). A Cy5-labeled, single-cysteine (Cys) L1 mutant, L1(T202C), was prepared as previously described, with an approximately 65% labeling efficiency (8). An L9 single-Cys mutant, L9(Q18C), was designed using multiple sequence alignments from a variety of bacterial strains to identify poorly-conserved L9 amino acid residues in combination with X-ray crystallographic (9, 10) and cryo-EM structures (11, 12) of ribosomal complexes to identify L9 amino acid residues within FRET distance of our labeling position on L1(T202C). L9(Q18C) was constructed from the pProEX-HTb plasmid bearing the cloned, wild-type L9 gene using the QuickChange Mutagenesis System (Stratagene, Inc.) and verified by DNA sequencing. L9(Q18C) was overexpressed and purified using $Ni^{2+}$-nitrilotriacetic acid affinity chromatography (Qiagen) under denaturing buffer conditions specified by the manufacturer. Purified L9(Q18C) was then renatured in Renaturation Buffer (50 mM sodium phosphate (pH=7.2) and 100 mM NaCl). The 6xHis tag was subsequently cleaved by incubating L9(Q18C) with TeV protease at 4 $^{\circ}$C overnight in Renaturation Buffer. Cleaved L9(Q18C) was purified from the cleaved 6xHis tag, uncleaved L9(Q18C), and TeV protease using a second $Ni^{2+}$-nitrilotriacetic acid affinity chromatography step in Renaturation Buffer. Fluorescent labeling of L9(Q18C) with Cy3-maleimide (GE Lifesciences) was performed by incubating 40 μM L9(Q18C) and 800 μM Cy3-maleimide at room temperature for two hours in a buffer containing 50 mM Tris-HCl ($pH_{25\,^{\circ}C}$ = 7.0), 200 mM KCl, 4 mM tris(2-carboxyethyl)phosphine (TCEP) and 4 M urea. Cy3-labeled L9(Q18C) was purified from free, unreacted dye by gel filtration on Superdex 75 in a buffer containing 20 mM Tris-HCl ($pH_{4\,^{\circ}C}$ = 7.5), 400 mM $NH_4Cl$, 4 mM $MgCl_2$ and 4 M urea. Buffers for purifying, labeling, and storing L9(Q18C) contained urea in order to prevent the aggregation and precipitation of L9(Q18C) which is observed in the absence of its ribosomal binding partner. Based on a comparison of Cy3 and L9(Q18C) concentrations determined from Cy3 absorbance at 550 nm (extinction coefficient = 150,000 $M^{-1}$ $cm^{-1}$) and an L9(Q18C) Bradford assay, we estimate ~50% labeling of L9(Q18C).

**Preparation of dual-labeled 50S subunits.** Cy5-labeled L1(T202C) and Cy3-labeled L9(Q18C) were reconstituted into purified 50S ribosomal subunits lacking L1 and L9 using previously described protocols (13, 14). Reconstituted, dual-labeled 50S subunits were subjected to sucrose density gradient ultracentrifugation in order to separate free, unincorporated, Cy5-labeled L1(T202C) and Cy3-labled L9(Q18C) from dual-labeled 50S subunits. Based on spectrophotometrically-determined 50S subunit, Cy5, and Cy3 concentrations, we estimate a reconstitution efficiency of approximately 100% for Cy5-labeled L1(T202C) and approximately 60% for Cy3-labeled L9(Q18C). Given labeling efficiencies of ~65% and ~50% for L1(T202C) and L9(Q18C), respectively, dual-labeled 50S subunits are estimated to contain ~65% Cy5-labeled L1(T202C) and ~30% Cy3-labeled L9(Q18C). However, only those 50S subunits carrying both Cy5-labeled L1(T202C) and Cy3-labeled L9(Q18C) will generate an observable smFRET$_{L1-L9}$ signal in our experiments. Ribosomes lacking Cy5-labled L1(T202C) and/or Cy3-labeled L9(Q18C) or harboring unlabeled L1(T202C) and/or L9(Q18C) are not detected and do not affect either the collected smFRET$_{L1-L9}$ data or its analysis.

***Biochemical characterization of dual-labeled 50S subunits.*** A standard primer-extension inhibition, or toeprinting, assay (15, 16) was used to test the ability of dual-labeled 50S subunits, 30S subunits, and fMet-tRNA$^{fMet}$ to properly initiate on a defined mRNA in the presence of initiation factors 1, 2, and 3 and GTP. Toeprinting was also used to verify that these ribosomal initiation complexes could undergo peptide-bond formation and translocation through two rounds of translation elongation. The results shown in Fig. S1 demonstrate that ribosomes harboring dual-labeled 50S subunits can undergo all of these reactions with an efficiency that is indistinguishable from that observed for ribosomes harboring wild-type 50S subunits. Based on these toeprinting results, we estimate that, upon addition of EF-Tu(GTP)Phe-tRNA$^{Phe}$ and EF-G, INI is ~90% active through the first round of translation elongation necessary to generate POST$_{fM/F}$ and, upon further addition of EF-Tu(GTP)Lys-tRNA$^{Lys}$ and EF-G, POST$_{fM/F}$ is ~70% active in the second round of translation elongation necessary to generate POST$_{F/K}$.

***Purification and surface immobilization of ribosomal complexes.*** INI and POST$_{fM/F}$ complexes for smFRET$_{L1-L9}$ studies were enzymatically prepared on our 5'-biotinylated, T4 gene product 32-derived mRNA using dual-labeled 50S subunits, 30S subunits, and all necessary initiation and/or elongation factors and aminoacyl-tRNAs. Likewise, ribosomal complexes analogous to INI and POST$_{fM/F}$ complexes but containing Cy5-labeled 50S subunits and Cy3-labeled P-site tRNAs for smFRET$_{L1-tRNA}$ experiments were enzymatically prepared on the same mRNA using 50S subunits harboring a Cy5-labeled L1 protein, 30S subunits, and all necessary initiation and/or elongation factors and aminoacyl-tRNAs, including either fMet-(Cy3)tRNA$^{fMet}$ or Phe-(Cy3)tRNA$^{Phe}$ (8). The resulting complexes were separated from free mRNA, translation factors, and aminoacyl-tRNAs by sucrose density gradient ultracentrifugation as previously described (1, 8). Purified ribosomal complexes were immobilized *via* a biotin-streptavidin interaction onto the surface of a streptavidin-derivatized quartz flow cell. As previously reported, prior to immobilization of ribosomal complexes, quartz flow cells were passivated by amino silanization followed by reaction with a mixture of *N*-hydroxysuccinimide ester-activated polyethyleneglycol (PEG) and PEG-biotin. Passivated flow cells were incubated with streptavidin just prior to use (1, 8).

***Total internal reflection fluorescence microscopy.*** We have designed and constructed a home-built, wide-field, prism-based total internal reflection fluorescence microscope utilizing a 532 nm laser (CrystaLaser) as an excitation source and a 512x512 pixel, back-thinned CCD camera (Cascade II, Princeton Instruments) as a detector. This microscope allows direct visualization of approximately 200-300 ribosomal complexes in an observation area of 60x120 $\mu m^2$. All smFRET$_{L1-L9}$ data were collected under 11mW excitation laser power with 0.10 sec frame$^{-1}$ time resolution; all smFRET$_{L1-tRNA}$ data were collected under 15mW excitation laser power with 0.05 sec frame$^{-1}$ time resolution. Single ribosomes were identified by single-step fluorophore photobleaching.

smFRET$_{L1-L9}$ data were recorded using a 0.10 sec frame$^{-1}$ time resolution in order to maximize the signal-to-noise such that transitions between the relatively closely spaced 0.34 and 0.56 FRET states (corresponding to a 0.22 FRET difference) could be easily identified and analyzed. The smFRET$_{L1-tRNA}$ data were recorded using a 0.05 sec frame$^{-1}$ time resolution in order to remain consistent with, and allow direct comparison to, our previously reported smFRET$_{L1-tRNA}$ data (8) Transitions between the relatively further spaced 0.21 and 0.84 FRET states (corresponding to a 0.63 FRET difference) for the smFRET$_{L1-tRNA}$ data recorded using Cy3-labeled tRNA$^{Phe}$ at the P-site (8) or 0.10 and 0.60 FRET states (corresponding to a 0.50 FRET difference) for the smFRET$_{L1-tRNA}$ data recorded using Cy3-labeled tRNA$^{fMet}$ at the P-site (the current work) are easily observed regardless of the slightly lower signal-to-noise in the higher time resolution data.

***Selection of smFRET vs. time trajectories.*** Raw fluorescence intensity data were analyzed with the MetaMorph software suite (Molecular Devices). Selection of smFRET trajectories was performed as previously described (8). Due to imperfect performance of emission filters, which allow a small amount of Cy3 emission to bleed through into the Cy5 emission channel, the Cy5 intensity of each trace was corrected using bleed-through coefficient of 7% (experimentally measured using Cy3-labeled DNA oligonucleotides). Both Cy3 and Cy5 intensities were baseline corrected such that the averaged post-photobleaching intensity for both fluorophores is centered at zero intensity. FRET values were calculated using $I_{Cy5}/(I_{Cy3}+I_{Cy5})$, where $I_{Cy3}$ and $I_{Cy5}$ are the emission intensities for Cy3 and Cy5, respectively, for each baseline-corrected pair of Cy3 and Cy5 data points in each trace (8).

***Dwell Time Analysis.*** Each smFRET trajectory was idealized as a hidden Markov model, using the vbFRET software package ((17)– open source MATLAB code to be available at vbfret.sourceforge.net upon acceptance of manuscript for publication). vbFRET infers the idealized trajectory using a variational Bayesian analysis (rather than maximum likelihood) (18), which also determines kinetic parameters as well as the number of conformational states for individual trajectories, thereby avoiding overfitting. Although rare, transitions in the idealized smFRET trajectories occurring with a change of less than 0.05 FRET or 0.1 FRET were discarded from the analysis of smFRET$_{L1-L9}$ or smFRET$_{L1-tRNA}$ datasets, respectively. For each data set, the data points from the entire set of idealized smFRET trajectories were used to generate a one-dimensional FRET histogram. Origin7.0 was used to fit each histogram with three Gaussian distributions using initial guesses centered at 0, 0.35, and 0.55 for smFRET$_{L1-L9}$ and centered at 0, 0.10 and 0.65 for smFRET$_{L1-tRNA}$. Thresholds for each FRET state were set using the full width at half height of the Gaussian distribution. Using these thresholds, the dwell time in each state prior to transitioning was extracted from the idealized smFRET trajectories. One-dimensional population *vs.* time histograms were plotted and lifetimes were determined by fitting the histogram to either a single- or double-exponential decay (Fig. S2). Transition rates were calculated by taking the inverse of the lifetimes and applying corrections for premature truncation due to photobleaching as well as the finite nature of the trajectory. (refs. (19, 20) and Table S1).

***Subpopulation and dwell time analysis of PMN$_{F/-}$ + EF-G(GDPNP).*** Binding of EF-G(GDPNP) to PMN$_{F/-}$ strongly inhibits closed→open L1 stalk transitions such that the open⇄closed L1 stalk equilibrium shifts to favor the closed L1 stalk conformation and the rate of photobleaching from the closed L1 stalk conformation effectively out competes closed→open transitions; the overall effect is a decrease in the occupancy of SP$_{fluct}$ and a corresponding increase in the occupancy of SP$_{closed}$. Because EF-G(GDPNP)-bound PMN$_{F/-}$ occupies SP$_{closed}$, $k_{open}$ and $k_{close}$ for EF-G(GDPNP)-bound PMN$_{F/-}$ cannot be calculated (Table 1). Despite this, Fig. 2B reveals that 23% of the PMN$_{F/-}$ + EF-G(GDPNP) trajectories remain in SP$_{fluct}$. Analysis of SP$_{fluct}$ in this scenario is complicated by the compositional heterogeneity present in PMN$_{F/-}$ + EF-G(GDPNP) that arises from incomplete reactivity at each of the various enzymatic steps required to prepare PMN$_{F/-}$ + EF-G(GDPNP) (i.e. reaction of INI with EF-Tu(GTP)Phe-tRNA$^{Phe}$ to generate PRE$_{fM/F}$, reaction of PRE$_{fM/F}$ with EF-G to form POST$_{fM/F}$, puromycin reaction of POST$_{fM/F}$ to form PMN$_{F/-}$, and binding of EF-G(GDPNP) to PMN$_{F/-}$). Of these potential sources of heterogeneity, the ones that primarily contribute to SP$_{fluct}$ in PMN$_{F/-}$ + EF-G(GDPNP) are: (1) binding of EF-G(GDPNP) to a residual amount (estimated at ~10% (Fig. S1)) of PMN$_{fM/-}$ that arises from puromycin reaction of INI that failed to undergo elongation. This source of heterogeneity can be easily resolved by investigating EF-G(GDPNP)-bound PMN$_{fM/-}$ (Tables 1 and S2); (2) PRE$_{F/-}$ that failed to bind EF-G(GDPNP). This source of heterogeneity can be resolved using the dwell time analysis of PMN$_{F/-}$ in the absence of EF-G(GDPNP) (Tables 1 and S2). All other potential sources of compositional heterogeneity (i.e. PMN$_{fM/-}$ that failed to bind EF-G(GDPNP) or INI or POST$_{fM/F}$ that failed to undergo puromycin reaction) are either negligible or result in trajectories that primarily occupy SP$_{open}$, thus not affecting the dwell time analysis of SP$_{fluct}$.

Based on the sources of heterogeneity described above, the dwell time histogram for the open L1 stalk conformation of PMN$_{F/-}$ in the presence of 1 μM EF-G(GDPNP) was fit with a double-exponential decay in which A$_1$ represents the relative population of transition events contributed by PMN$_{F/-}$ that failed to bind EF-G(GDPNP) and A$_2$ represents the relative population of transition events contributed by EF-G(GDPNP)-bound PRE$_{fM/-}$. In order to convert the relative populations of transition events, A$_1$ and A$_2$, into relative populations of trajectories, P$_1$ and P$_2$, we need to account for the fact that the fast-transitioning population of trajectories, P$_2$, will have a larger contribution to the total number of transition events than the slow-transitioning trajectory population, P$_1$. Based on the A$_1$ and A$_2$ lifetimes, 1.5 sec and 0.35 sec, respectively, the ratio of "transition frequencies" for P$_1$ and P$_2$ can be estimated as 0.35:1.5. Thus, solving the following equation: (0.35P$_1$)/(1.5P$_2$) = A$_1$/A$_2$ = 14/86, yields P$_1$/P$_2$ = 0.7. Therefore, (23%)[0.7/(1+0.7)] = 9% is the percentage of PRE$_{F/-}$ complexes that do not bind EF-G(GDPNP) and 23%-9% = 14% is the percentage of contaminating PRE$_{fM/-}$ complexes. These results are consistent with our toeprinting activity assays (Fig S1) and are further supported by dwell time analyses of PRE$_{F/-}$ as a function of EF-G(GDPNP) concentration, in which A$_1$ decreases and A$_2$ increases with increasing concentrations of EF-G(GDPNP) (Table S2).

**Supporting Table 1**

*Table S1.* Lifetimes of fluorophores prior to photobleaching from each FRET state[a].

| FRET state | Labeled Components | Lifetime (sec) |
|---|---|---|
| 0.10[b] | (Cy5)L1, (Cy3)tRNA$^{fMet}$ | 16.3±1.7 |
| 0.34[c] | (Cy5)L1, (Cy3)L9 | 8.0±1.4 |
| 0.56[d] | (Cy5)L1, (Cy3)L9 | 7.7±1.8 |
| 0.60[e] | (Cy5)L1, (Cy3)tRNA$^{Phe}$ | 4.0±0.4 |

[a] Lifetimes are the average values measured from the datasets in which the majority of the sample population stably samples the FRET state designated in the table.

[b] Lifetime of the 0.10 FRET state is extracted from a ribosomal initiation complex analogous to INI, but carrying the labeled components designated in the table.

[c] Lifetime of the 0.34 FRET state is extracted from PMN$_{F/-}$ in the presence of 500nM and 1µM of EF-G(GDPNP).

[d] Lifetime of the 0.56 FRET state is extracted from POST$_{fM/F}$, POST$_{-/F}$ and POST$_{-/K}$.

[e] Lifetime of the 0.60 FRET state is extracted from a PMN complex analogous to PMN$_{F/-}$ in the presence of 1µM EF-G(GDPNP), but carrying the labeled components designated in the table.

**Supporting Table 2**

*Table S2.* L1 stalk closing and opening rates for $PMN_{F/-}$ and $PMN_{fM/-}$ as a function of EF-G concentration[a].

| Complex | [EF-G][b] | $k_{close}$ (sec$^{-1}$) | | | | $k_{open}$ (sec$^{-1}$) |
|---|---|---|---|---|---|---|
| | | $k_1$ | $A_1$(%) | $k_2$ | $A_2$(%) | |
| $PMN_{F/-}$[c] | 0 nM | 0.51±0.06 | | | | 0.84±0.17 |
| | 5 nM | 0.62±0.08 | | | | 0.91±0.03 |
| | 50 nM | 0.44±0.21 | 61±9 | 2.7±0.3 | 39±9 | 1.2±0.4 |
| | 0.5 µM | 0.44±0.23 | 26±10 | 3.2±0.3 | 74±10 | 1.0±0.2 |
| | 1 µM | 0.51±0.03[d] | 14±8 | 2.7±0.7 | 86±8 | 1.0±0.2 |
| $PMN_{fM/-}$[e] | 0 nM | 0.37±0.09 | | | | 1.5±0.3 |
| | 5 nM | 0.25±0.09 | 29±4 | 2.6±0.7 | 71±4 | 1.4±0.1 |
| | 50 nM | 0.31±0.08 | 11±9 | 3.1±0.9 | 89±9 | 1.4±0.2 |
| | 0.5 µM | | | 2.5±0.6 | | 1.32±0.07 |
| | 1 µM | | | 3.0±0.6 | | 1.2±0.2 |

[a] All rates are derived from dwell time analysis of only those trajectories which occupy $SP_{fluct}$. Rates reported here are the average and standard deviation from three independent datasets. All rates were corrected for premature truncation due to fluorophore photobleaching as well as the finite nature of the smFRET trajectories (see Methods and Table S1).

[b] The concentration of GDPNP in all experiments was 1 mM.

[c] In this case, trajectories in $SP_{fluct}$ encompass residual amounts of EF-G(GDPNP)-free $PMN_{F/-}$ and EF-G(GDPNP)-bound $PMN_{fM/-}$ and thus the reported rates do not represent EF-G(GDPNP)-bound $PMN_{F/-}$. Instead, $A_1$ represents the relative population of transition events contributed by EF-G(GDPNP)-free $PMN_{F/-}$ and $A_2$ represents the relative population of transition events contributed by EF-G(GDPNP)-bound $PMN_{fM/-}$ (SI Methods).

[d] As a consequence of the compositional heterogeneity described in footnote (c), above, and in the SI Methods, the dwell time histogram for the open L1 stalk conformation was fitted to a double-exponential decay in which the lifetime associated with $A_1$, $\tau_1$, was set to 1.5 sec (i.e. the lifetime of the open L1 stalk conformation as experimentally measured for $PMN_{F/-}$ in the absence of EF-G(GDPNP)), yielding $k_1$ = 0.51 sec$^{-1}$.

[e] $A_1$ represents the relative population of transition events that are contributed by $PMN_{fM/-}$ that has failed to bind EF-G(GDPNP). $A_2$ represents the relative population of transition events contributed by EF-G(GDPNP)-bound $PMN_{fM/-}$.

# SUPPORTING FIGURES

*Figure S1.* **Primer-extension inhibition, or toeprinting, assay.** The activity of ribosomes harboring reconstituted, dual-labeled 50S subunits was tested using a primer-extension inhibition, or toeprinting, assay (15, 16). Translation reactions were performed using all purified components and an mRNA pre-annealed with a $^{32}$P-labeled DNA primer. The position of the initiated ribosomal complex on the mRNA was determined by monitoring the inhibition of a subsequent reverse transcription reaction and running the cDNA products of the reverse transcription reaction on a denaturing PAGE. cDNA bands corresponding to mRNA positions +15, +16, +18, and +21, relative to the A of the AUG start codon which comprises position 0, report on the initiated ribosomal complex (+15), the incorporation of the first A-site tRNA (Phe-tRNA$^{Phe}$) (+16), the first translocation step (+18), and, collectively, the incorporation of a second A-site tRNA (Lys-tRNA$^{Lys}$) and translocation step (+21). Lane 1 is a control generated by reverse transcription of the $^{32}$P-labeled primer-annealed mRNA in the absence of ribosomes; this control reports on intrinsic reverse transcriptase inhibition sites likely caused by regions of stable secondary structures within the mRNA. Raw intensities at mRNA nucleotide positions +15, +16, +18 and/or +21 in Lanes 2- 9 were therefore corrected using the Lane 1 intensities at the corresponding positions. Comparison of the corrected bands at +15, +18, and +21 in Lanes 7-9 suggest that initiated ribosomal complexes are ~90% active in the first round of elongation and ~70% active in the second round of elongation. These activities are indistinguishable from those of wild-type ribosomes (Lanes 3-5).

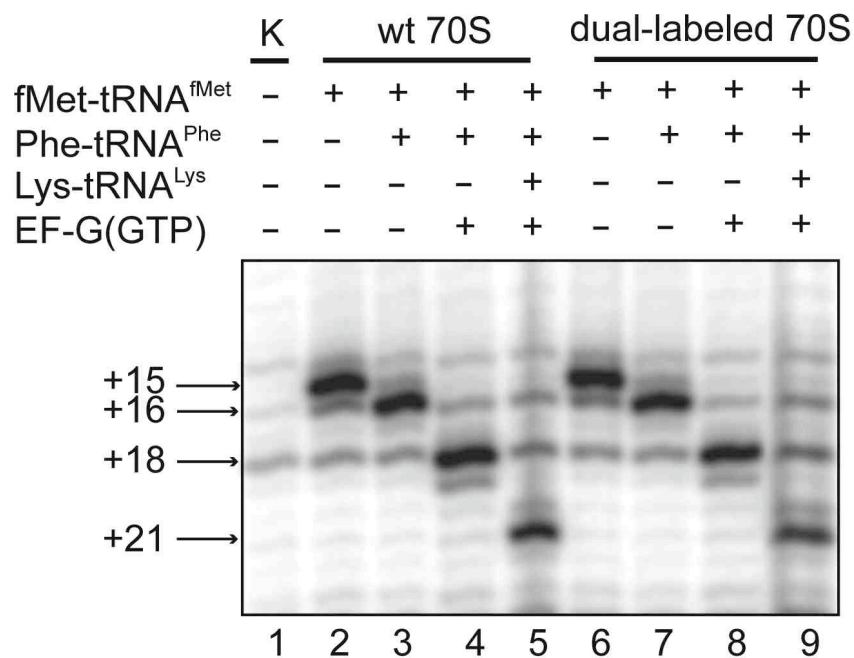

*Figure S2.* **Sample dwell time analysis.** (A) A transition density plot for each complex is generated by plotting the "Starting FRET" *versus* the "Ending FRET" for each transition as a surface contour plot of two-dimensional population histograms. Contours are plotted from tan (lowest population) to red (highest population). (B) One-dimensional FRET histograms calculated from the idealized smFRET trajectories generated by hidden Markov modeling of the raw smFRET trajectories using vbFRET (ref. 17, 18 --- open source MATLAB code available at vbfret.sourceforge.net). Initial thresholds for each FRET state were determined by fitting these histograms to three Gaussian distributions with user-specified initial guess values of 0, 0.35, and 0.55 FRET for the Gaussian centers and using the full width at half height of the resulting Gaussians as initial threshold values. (C) Dwell time histograms in the 0.56 FRET and 0.34 FRET states are described either by a single-exponential decay ($A*\exp(-t/t_0)+y_0$) or a double-exponential decay ($A_1*\exp(-t/t_1)+A_2*\exp(-t/t_2)+y_0$).

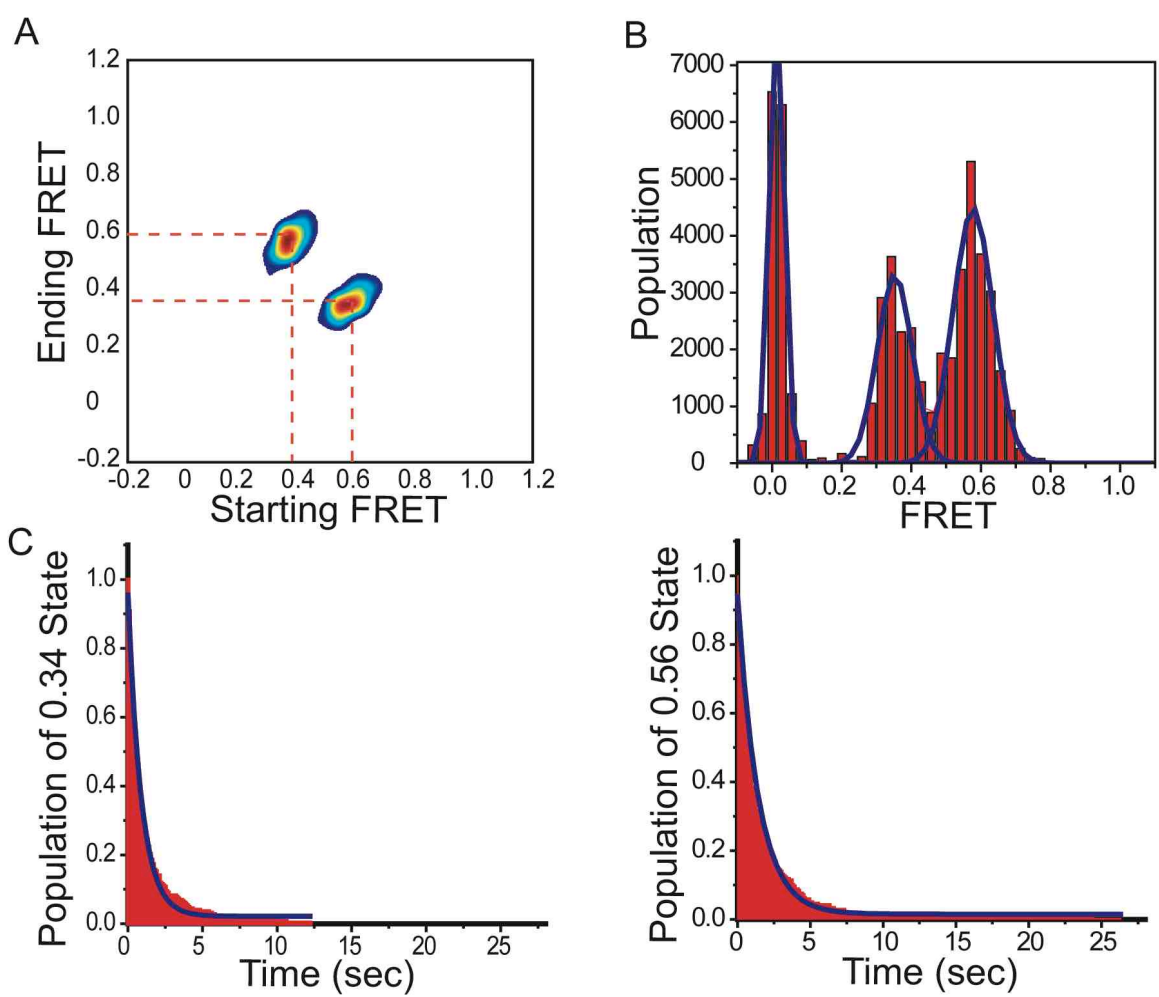

*Figure S3.* **Steady-state smFRET$_{L1\text{-tRNA}}$ of PRE and PMN complexes analogous to PRE$_{fM/F}$ and PMN$_{fM/-}$ in the absence and presence of 1 μM EF-G(GDPNP).** Cartoon representations of PRE and PMN complexes analogous to PRE$_{fM/F}$ and PMN$_{fM/-}$ depict the 30S and 50S subunits in tan and lavender, respectively, with the L1 stalk in dark blue, tRNA$^{fMet}$ as an orange line, EF-G in light purple, and Cy5 and Cy3 as red and green stars, respectively, (first row). Representative Cy3 and Cy5 emission intensities are shown in green and red, respectively (second row). The corresponding smFRET *vs.* time trajectories, in which the FRET efficiency is calculated using the equation $I_{Cy5}/(I_{Cy3}+I_{Cy5})$, where $I_{Cy3}$ and $I_{Cy5}$ are the emission intensities of Cy3 and Cy5, respectively, are shown in blue (third row). Surface contour plots of the time evolution of population FRET are plotted from tan (lowest population) to red (highest population) (bottom row). The number of traces that were used to construct each contour plot is indicated by "N". (A) PRE complex analogous to PRE$_{fM/F}$, generated by addition of 100 nM EF-Tu(GTP)Phe-tRNA$^{Phe}$ to a ribosomal initiation complex analogous to INI. (B) PMN complex analogous to PMN$_{fM/-}$, generated by addition of 1 mM puromycin to a ribosomal initiation complex analogous to INI, in the absence of EF-G(GDPNP). (C) PMN complex analogous to PMN$_{fM/-}$ in the presence of 1μM EF-G(GDPNP). The 0.10 and 0.60 FRET states here correspond to the 0.21 and 0.84 FRET states measured previously (8). The slightly lower FRET values reported in the present work are due to the different labeling position on tRNA$^{fMet}$ (s$^4$U8, this work) *vs.* tRNA$^{Phe}$ (acp$^3$U47, previous work (8)) as well as the use of a slightly different image-splitting device for separating the Cy3 and Cy5 emission wavelengths (Dual-View (Photometrics), this work *vs.* Quad-View (Photometrics), previous work (8). We note here the observation that the smFRET$_{L1\text{-tRNA}}$ signals for PRE$_{fM/F}$ and PMN$_{fM/-}$ exhibit poor signal-to-noise relative to PRE$_{F/K}$ and PMN$_{F/-}$; this is due to the higher noise in the Cy3 donor signal when Cy3 is covalently attached to the s$^4$U8 position of tRNA$^{fMet}$ *vs.* the acp$^3$U47 position of tRNA$^{Phe}$.

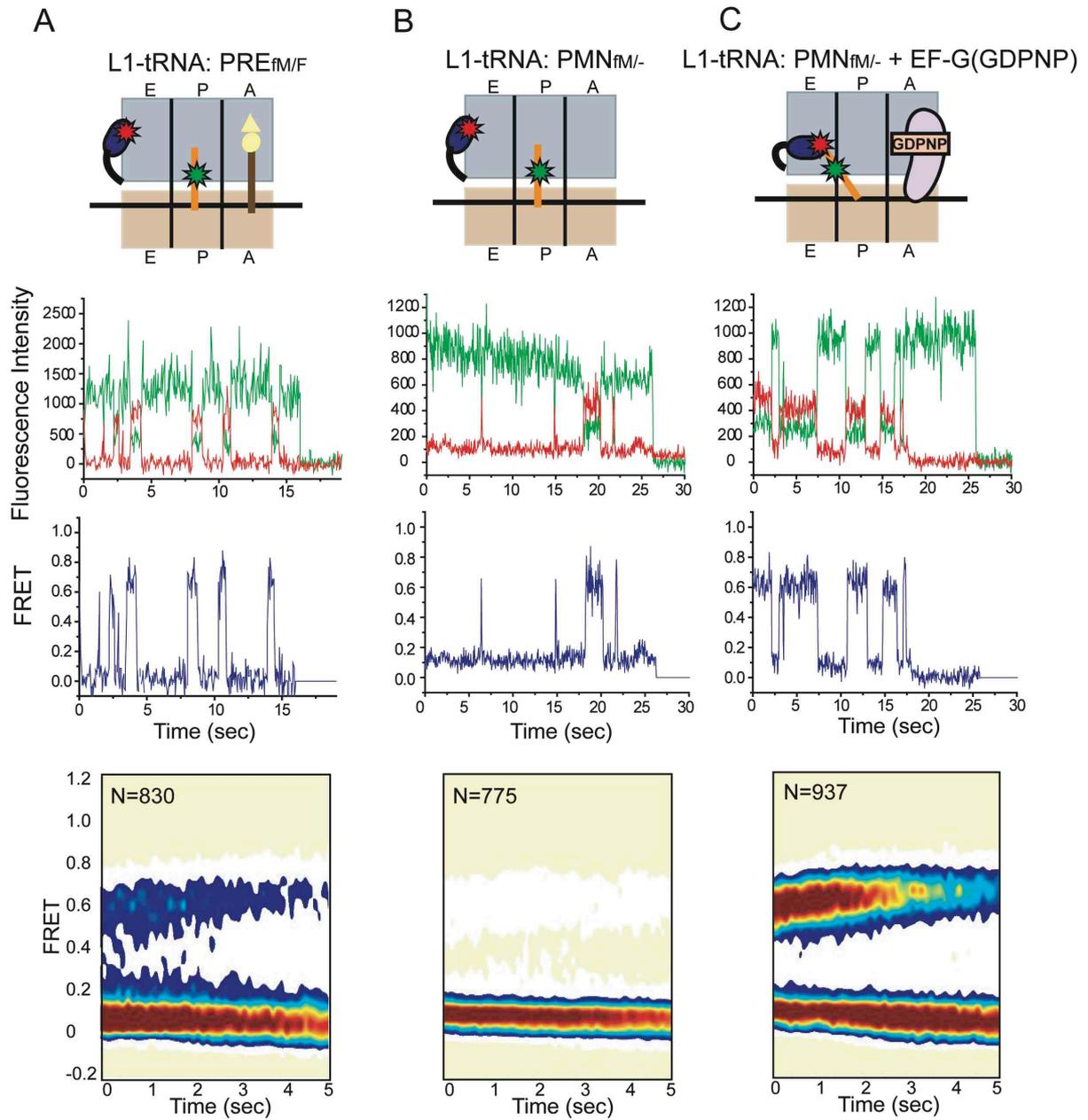

*Figure S4.* **L1 stalk dynamics as a function of EF-G concentration.** (A) From left to right, surface contour plots of the time evolution of population FRET for $PMN_{F/-}$ in the presence of 0 nM, 5 nM, 50 nM, 500 nM and 1 μM EF-G; the GDPNP concentration was 1 mM at each EF-G concentration. (B) From left to right, surface contour plots of the time evolution of population FRET for $PMN_{fM/-}$ in the presence of 0 nM, 5 nM, 50 nM, 500 nM and 1 uM of EF-G; the GDPNP concentration was 1 mM at each EF-G concentration. (C) Bar graph reporting the occupancies of $SP_{open}$, $SP_{closed}$, and $SP_{fluct}$ as a function of EF-G concentration for $PMN_{F/-}$ (left) and $PMN_{fM/-}$ (right).

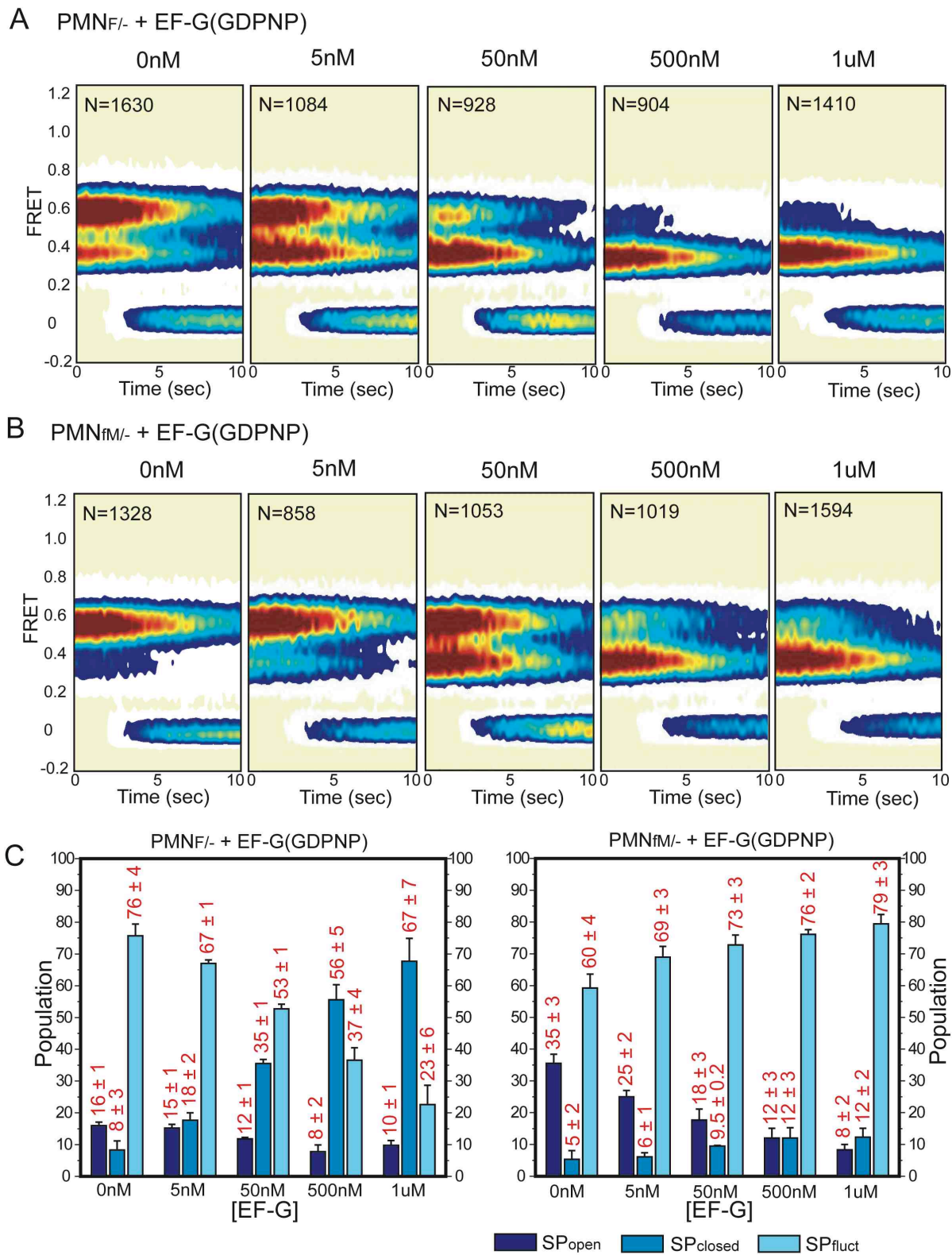

*Figure S5.* **Single-molecule E-site tRNA release assay.** Ribosomal complexes analogous to INI and POST$_{fM/F}$, formed using 50S subunits harboring a Cy5-labeled L1 and carrying fMet-(Cy3)tRNA$^{fMet}$ or fMetPhe-(Cy3)tRNA$^{Phe}$ at the P site, respectively, were immobilized *via* a biotinylated mRNA onto the surface of a streptavidin-derivatized quartz flow cell. Spatially-localized Cy3 fluorescence from individual surface-immobilized complexes was recorded as a function of time. Stopped-flow delivery of 100 nM EF-Tu(GTP)Phe-tRNA$^{Phe}$ in the presence of 1 μM EF-G(GTP) to initiation complexes resulted in peptide bond formation and translocation of the mRNA-tRNA complex. The translocation event placed the newly deacylated OH-(Cy3)tRNA$^{Phe}$ into the E site. Single OH-(Cy3)tRNA$^{fMet}$ dissociation events from the E site were followed in real-time by monitoring the loss of spatially-localized Cy3 signals (green triangles). In order to reduce the contribution of fluorophore photbleaching to the loss of spatially-localized Cy3 signals the 532 nm excitation laser was shuttered at 12 frames min$^{-1}$ for the first 5 frames, 6 frames min$^{-1}$ for the next 5 frames and 2 frames min$^{-1}$ for the last 30 frames. Similarly, release of (Cy3)tRNA$^{Phe}$ is triggered by stopped-flow delivery of 100 nM EF-Tu(GTP)Lys-tRNA$^{Lys}$ in the presence of 1 μM EF-G(GTP) to a POST complex analogous to POST$_{fM/F/-}$ (blue triangles). As a control, the intrinsic loss of spatially-localized Cy3 signals due to photobleaching and ribosome dissociation from the surface for both tRNA$^{fMet}$ (black squares) and tRNA$^{Phe}$ (purple circles) were recorded using identical shuttering parameters. The E-site tRNA release data were best described by double exponential decays (red curves) of the form $A_1*\exp(-t/t_1)+A_2*\exp(-t/t_2)+y_0$ for both tRNA$^{fMet}$ and tRNA$^{Phe}$. The relative populations and lifetimes of the slow and fast dissociating components are reported as the average value taken from three independent measurements. For tRNA$^{fMet}$, (33±3)% of the population exhibited a lifetime of 20±6 sec and (67±3)% of the population exhibited a lifetime of 430±30 sec, with measurement of the actual dissociation time limited by the photobleaching rate of Cy3. For tRNA$^{Phe}$, (37±9)% of the population exhibited a lifetime of 20±3 sec and (63±9)% of the population exhibited a lifetime of 310±50 sec.

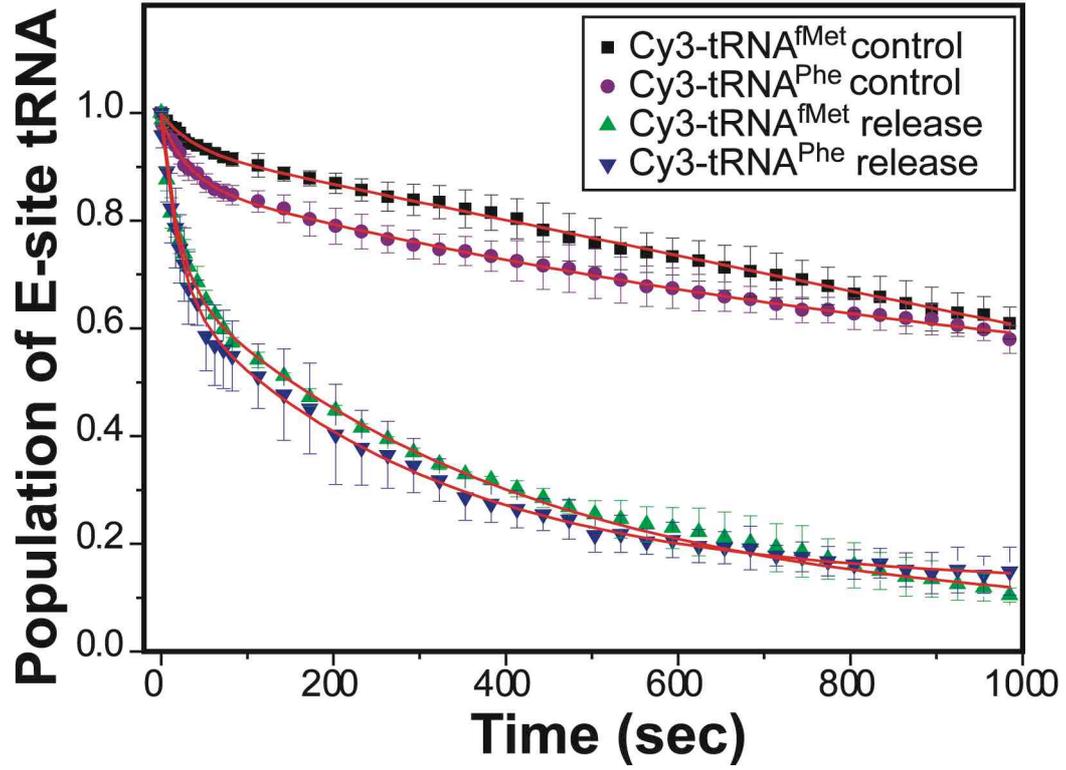

*Figure S6.* **The effect of the E-site tRNA on L1 stalk dynamics within POST complexes.** (A) Bar plot of the ccupancies of $SP_{open}$, $SP_{closed}$, and $SP_{fluct}$ in $POST_{-/F}$, $POST_{-/K}$, $POST_{fM*/F}$ and $POST_{F*/K}$. The means and standard deviations, calculated from three independent datasets, are shown as red numbers. (B) Cartoon representation and surface contour plot of the time evolution of population FRET for $POST_{-/F}$, prepared by the addition of 100 nM EF-Tu(GTP)Phe-tRNA$^{Phe}$ and 1 μM EF-G(GTP) to INI in non-polyamine buffer, followed by incubation at room temperature for 5 minutes and buffer exchange into Tris-Polymix buffer just prior to data collection. (C) Cartoon representation and surface contour plot of the time evolution of population FRET for $POST_{-/K}$, prepared by the addition of 100 nM EF-Tu(GTP)Lys-tRNA$^{Lys}$ and 1 μM EF-G(GTP) to $POST_{fM/F}$ in non-polyamine buffer, followed by incubation at room temperature for 5 minutes and buffer exchange into Tris-Polymix buffer just prior to data collection. (D) Cartoon representation and surface contour plots of the time evolution of population FRET for $POST_{fM*/F}$, prepared by addition of 100 nM EF-Tu(GTP)Phe-tRNA$^{Phe}$ and 1 μM EF-G(GTP) to INI in non-polyamine buffer, followed by incubation at room temperature for 5 minutes and buffer exchange into Tris-Polymix buffer, supplemented with 1 μM deacylated tRNA$^{fMet}$, just prior to data collection. (E) Cartoon representation and surface contour plots of the time evolution of population FRET for $POST_{F*/K}$, prepared by addition of 100 nM EF-Tu(GTP)Lys-tRNA$^{Lys}$ and 1 μM EF-G(GTP) to $POST_{fM/F}$ in non-polyamine buffer, followed by incubation at room temperature for 5 minutes and buffer exchange into Tris-Polymix buffer, supplemented with 1 μM deacylated tRNA$^{Phe}$, just prior to data collection.

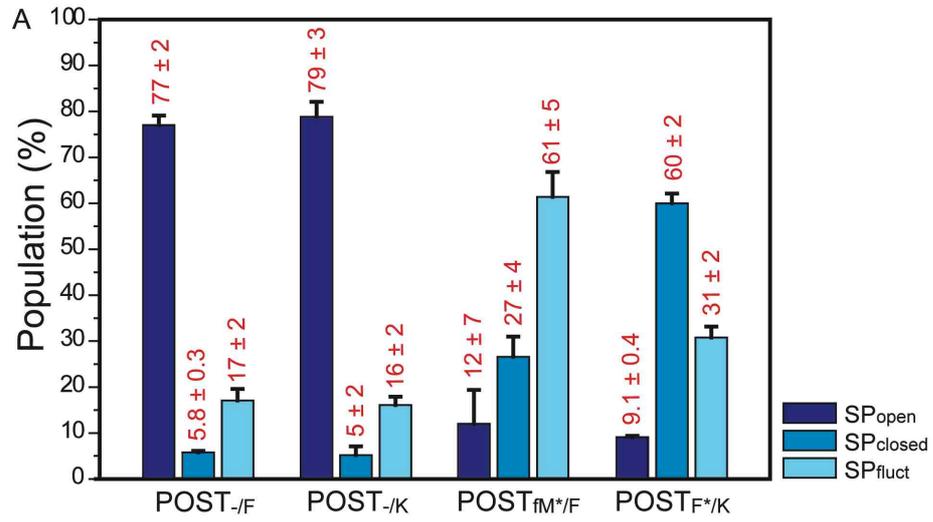
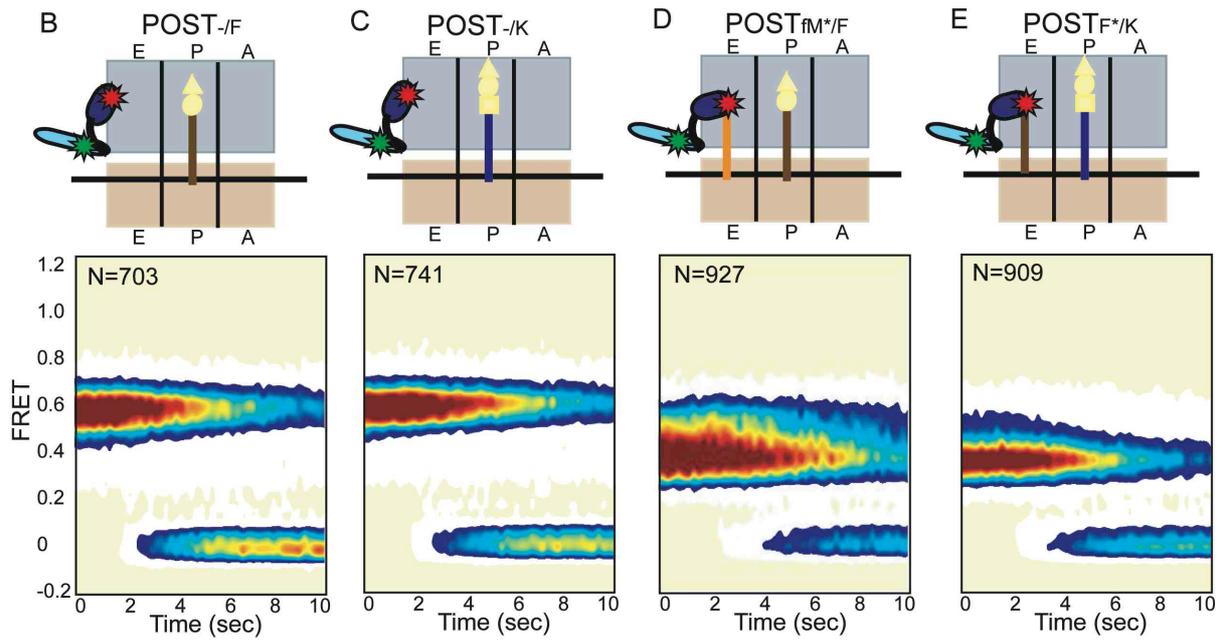